\def\mnras{MNRAS}
\def\apj{ApJ}
\def\nature{Nature}
\def\aa{{A \& A}}

\def\a{\alpha}
\def\b{\beta}
\def\g{\gamma}

\def\s{\sigma}

\def\e{\epsilon}
\def\z{\zeta}

\def\u{\delta}

\def\m{\mu}

\documentstyle[11pt]{article}
\begin{document}


\section*{Effects of Shocks on Emission from AGN's Central Engines -- I}


\centerline{R.Sivron, D.Caditz and S.Tsuruta}

\centerline{Physics Department, Montana State University, Bozeman, MT 59717}

\centerline{Email: sivron@physics.montana.edu}


\begin{abstract}
In this paper we show that
perturbations of the accretion flow within the central
engines of  some active galactic nuclei
(AGN) are likely to form shock waves in the accreting plasma.
Such shocks, which may be either collisional or collisionless, can
contribute to the observed high energy temporal and spectral 
variability. Our rationale is the
following: Observations show that the continuum emission probably
originates in an optically thin, hot plasma in the AGN central engine. 
The flux and spectrum from this hot plasma varies significantly 
over light-crossing timescales. Several authors have suggested that
macroscopic perturbations contained within this plasma are the
sources of this variability. In order to produce the observed
emission the perturbations must be radiatively coupled with the
optically thin hot matter and must also move with high velocities.
We suggest that shocks, which can be very
effective in randomizing the bulk motion of the perturbations,
are responsible for this coupling.
Shocks {\it should} form
in the central engine, because the temperatures and magnetic fields 
are probably reduced
below their virial values by radiative dissipation. Perturbations
moving at Keplerian speeds, or strong nonlinear excitations,
result in  supersonic and superAlf\'venic velocities leading to
shock waves within the hot plasma.

We show that even a perturbation smaller than the emitting region
can form a shock which significantly modifies the continuum emission
in an AGN, and that the spectral and
temporal variability from such a shock 
generally resembles those of radio quiet AGN.
As an example, the shock inducing  perturbation
in our model is a
small main sequence star, the capturing and eventual accretion of which
are known to be a plausible process. 
We argue that shocks in the central engine may also provide a natural 
triggering mechanism for the `cold' component of Guilbert \& Rees 
two-phase medium and an efficient mechanism for angular momentum transfer.
Current and future missions, such as Asca, XTE, XMM, AXAF and ASTRO-E may 
determine the importance
of shock related emission from the central engines of AGN.

\end{abstract}

%

\section{Introduction}

Recent observations indicate that
the continuum emission from AGN originates within
a hot plasma (e.g., Mushotzky, 
Done \& Pounds 1993, hereafter MDP93, and references therein).
Variability arguments suggest that the plasma
is contained within a region of size $R_X \sim t_d c$, 
where $t_d$ is the flux doubling
timescale. Within the commonly accepted accretion model for
AGN, the hot plasma is probably located in the vicinity of a 
central supermassive black hole 
(Rees 1984 and references therein; Blandford 1990, hereafter B90).
In this model, most of the continuum emission originates from within
about twenty gravitational radii of the black hole in the
region generally referred to as the AGN central engine.

An extensively considered accretion model of AGN, the
simple optically thick geometrically thin disk model
which does not include hot plasma, does not adequately explain
the continuum UV to $\gamma$ ray spectrum from AGN.
Several
modifications to the thin disk geometry have been proposed:
(i) the geometrically thick accretion disk supported by radiation pressure
(Abramowicz, Calvani \& Nobili 1980) with
a geometry which is susceptible to global
perturbations (Papaloizou and pringle 1984),
(ii) the geometrically thick, optically thin disk model with 
two temperature relativistic plasmas, first suggested for an X-ray binary
(Shapiro Lightman \& Eardley 1976) but which has been
extended to AGN counterparts (e.g., Rees et al. 1982; 
White \& Lightman 1989, hereafter  WL89; Tritz and Tsuruta 1989; 
and Narayan \& Li 1994), and (iii) a disk-corona model
(Liang \& Price 1977; Haardt \& Maraschi 1991; \.Zycki, 
Collin-Soufrin \& Czerny 1995, hereafter ZCC95; and Tsuruta
and Kellen 1995, hereafter TK95). 
For the two-temperature torus model
Narayan \& Li (1994) propose that at lower accretion 
rates advection lowers the efficiency of the accretion process and 
stabilizes the otherwise unstable hot electrons, which
suggests that this model may be more appropriate for 
sub-Eddington sources (see also Artemova et. al. 1996).
A common feature of all models that attempt to reconstruct the 
higher energy ($\sim 1$ keV -- $1$MeV) power-law
spectrum is the introduction
of an optically thin hot plasma (with the electron
temperature $T_e> 10^8$K) which may dominate the emission from 
the central engine. 

Large amplitude temporal variability of the hot plasma emission 
is present in at least 50\% of
radio quiet AGN, and significant spectral variability of 
similar timescales is also very common
(MDP93 and references therein,
Green 1993, McHardy 1988). 
The temporal and spectral variability has been 
generally explained in the context of perturbations 
on the thin, cold disk model, but rarely in the 
context of the geometrical models mentioned above.
Models for the variability of radio quiet AGN 
include, for example,
the hot spots model (Abramowicz 1991; Wiita 1993, hereafter W93), 
where  a multitude of randomly distributed flares or `hot spots' 
embedded within a thin disk are responsible for the total variability.
Other models also assume randomly distributed perturbations on the thin disk 
(Pudritz \& Fahlman 1982, DeVries \&
Kuipers 1989). The temperature of the thin disk in the central engine 
is found to be of order $T \sim 10^6$K giving a sound speed of order $c_s
\sim 10^{-2}c$ or less (Frank, King \& Raine 1992, hereafter FKR92).
The total volume affected by {\it {a single}} perturbation is therefore 
too small
to easily account for the doubling of the entire continuum luminosity, making 
these models dependent on the existence of {\it {many}}
perturbations of characteristic size, time and brightness occurring
on the disk surface.
In contrast, within the hot plasma region, a disturbance
may propagate at a much greater speeds, affecting a larger volume of the
continuum emitting material. A single perturbation effectively coupled to
the hot plasma may transfer enough energy to the radiating region within 
typical flux doubling timescales to account for the observed 
continuum variability.

Shocks have been investigated in the context of radio loud AGN
with shocks propagating down a jet and interacting with
irregularities in the jet material (Qian et al. 1991), or other forms
of shocks (W93).
In this paper we investigate the possibility that a strong
perturbation, moving at supersonic or superAlf\'venic speeds 
in the hot plasma of the AGN central engine, forms a shock which modifies
the continuum emission (see also Sivron \& Tsuruta 1994).
Such shocks may effectively
couple macroscopic perturbations with the optically thin hot 
matter, modifying the continuum spectrum. Shocks are also 
effective in randomizing the bulk motion of the perturbations 
which may have important consequences for the 
outward transfer of angular momentum.

In our work, we utilize recent studies of the effect of pairs 
on the central engine structure
(Guilbert \& Rees 1988, hereafter GR88;
Lightman \& White 1988, hereafter LW88; Coppi \& Blandford 1990,
hereafter CB90; Sivron \& Tsuruta 1993, hereafter ST93; and
Ghisellini \& Haardt 1994, hereafter GH94).
These authors suggest that in
compact central engines, dampening processes 
readily cool a significant fraction of the hot matter,
radiating the excess energy away. The cooled matter then forms 
a cold `phase' component. We show that in central engines with
near-Eddington luminosity the
post-shock matter is, in effect, an `increased compactness' region in
which the shocked matter is cooled.
The shocks are thus a natural mechanism by which radiative power is
increased, and cold phase matter is created.
The cooling of post shock matter results in radiation
from the shock front which may be responsible for the `soft' flares, whereas
cold phase matter behind the shock front may be responsible
for the relatively hard `dips' in the light-curve (see section 4).

For mathematical simplicity
we use a model in which the source of the perturbations is local, for
example, a small main sequence star.
Such a star can be captured, and eventually accreted into the
central engine (Syer, Clarke \& Rees 1990,
hereafter referred to as SCR90). For a wide range of accretion parameters,
such an event naturally results in the creation of shocks. Such shocks
can significantly modify the light-curve from
the X-ray emitting region of AGN.

In section 2
we show that  macroscopic
inhomogeneities, which we assume travel at Keplarian speeds,
naturally posses supersonic or superAlf\'venic speeds when present in
the hot plasma of the AGN central engine for the three models discussed above. 
In section 3 we show that the shocks effectively couple the perturbations
with the hot plasma, and large enough
perturbations can therefore result in the modification of the total emission. 
In section 4 we calculate the effects of a large shock
on the light-curve of a model AGN. The main purpose of this exercise is to 
demonstrate the type of spectral variability expected when shocks
modify the emission from the central continuum source.
We discuss our results in section 5, and present our conclusions in section 6.


\section{Shock Formation}

Recent observations as well as theoretical analyses suggest that
hot plasma must exist in the central engine.
From an observational point of view,
the variable continuum of nearly power-law shape in the X-ray 
to $\gamma$-ray bands could be explained as emission from 
hot plasma radiating synchrotron self-Compton or Comptonized 
bremsstrahlung radiation from an accretion disk, filaments, or clouds 
(B90, ST93, GH94 and TK95). Theory predicts that the
simple thin accretion disk will be disrupted within 
a critical radius where heating due to
gravitational and viscous energy release surpasses the cooling by 
bremsstrahlung, the temperature rises and the gas pressure 
disrupts the thin disk (Rees et al. 1982, B90).
Other plasma instabilities may also disrupt the thin disk (see,
for example, W93).
The accretion flow inside this radius is not a simple thin disk and 
it may form a geometrically thick torus (cases (i) and 
(ii) above), or a disk-corona structure with dominant radiation
coming from corona (Haardt \& Maraschi 1991).

In order to describe the physical conditions within the central engine 
in the above models we use the following parameters:
The dimensionless radial distance to the supermassive black hole is 
$r = R / R_g$ where $R_g = GM/c^2$ is the gravitational radius. 
The bulk of X-ray continuum is thought to originate from within
$R_X \sim 20 R_g$.
The central black hole  mass, $M$, for 
Seyfert galaxies and radio quiet quasars has
been estimated from X-ray time variability data to 
be $M > \sim 10^6 M_{\odot}$ and we
therefore write the black hole mass in terms of $M_6 = M / 10^6 M_{\odot}$.
The accretion rate, $\dot M$, has been related to the X-ray luminosity, $L_X$,
through $\dot M = L_X / ( \eta c^2 )$,
where $\eta$ is the efficiency  of converting
gravitational energy of the accreted matter into radiation. Luminosities
of order $L_X \sim 10^{41-44}$ erg sec$^{-1}$ are typical for the
2 - 10 KeV band.
The dimensionless accretion rate is
${\dot m} = {\dot M} / {\dot M}_E$, where
\begin{equation}
{\dot M}_E= {4\pi G M m_p \over(\sigma_T c)},
\end{equation}
is the Eddington accretion rate,
$m_p$ is the proton mass and $\sigma_T$ the Thompson cross section.
Assuming $\eta\sim 0.1$ and $M_6 \sim 1$ we
find typical accretion rates of order $0.01 <\sim \dot m <\sim 10$.

\subsection{Collisional Shocks}

That collisional shocks form and Ranking-Hugoniot relations can be written,
when wave steepening occurs
in relativistic thermal plasmas was shown by Taub (1948).
Collision dominated waves and shocks may be present when
the particle deflection length is smaller than the size of a typical
large perturbation in the central engine, i.e. when:
\begin{equation}
\lambda \, \sim \, {2(kT_e)^2 \over \pi n_i e^4 \ln\Lambda } \, < \, R_g,
\end{equation}
where $n_i$ is the ion number density, $T_e$ the electron temperature, and 
the Coulomb logarithm is here taken to be $\ln\Lambda \sim 20$
(see e.g., FKR92).
Assuming quasi-spherical accretion geometry with solid angle $\Omega$,
we may easily relate the disk density to the accretion rate:
\begin{equation}
{\dot M} = {\Omega R^2 \mu m_p n_i v_{acc}}.
\end{equation}
Here $\mu m_p$ is the average ion mass and $v_{acc}$ is the radial
accretion velocity.
Equations (2) and (3) then give a
lower limit to the accretion rate for AGN which may support
collisional shocks:
\begin{equation}
{\dot m}\ge 0.08  \biggl(
{\Theta_e \over {\Theta_{v}}}\biggr)^2 \biggl({r\over
20}\biggr)^{-1/2}
\biggl({\Omega \over 4\pi}\biggr) \biggl({v_{acc}\over v_{ff}}\biggr),
\end{equation}
where $\Theta_e=kT_e/m_e c^2$ is the dimensionless electron temperature and
$\Theta_{v}=1/r$ is the dimensionless  virial temperature.
One can see that collision dominated waves may form in the 
near or super-Eddington  regime. This result is hardly surprising,
since it is related to that of GR88
which claim that the central engine is almost
always optically thick for near Eddington luminosity,
with the main difference being that the mean free path for photons in GR88
is replaced
by the deflection length for electron collisions.
We can compare the above result with the condition
in Rees et al (1982), ${\dot m}_{cr} \le 50 \a^2$
on the maximal accretion rate such that ions and
electrons decouple to form a two-temperature ion torus. 
Here $\a$ is similar to the
parameter in the standard $\a$ accretion model which relates
the stress and pressure in a thin disk. We see that even two temperature 
tori can marginally support collisional waves.

For the cases in which the shocks are collision dominated we
use the thermodynamic parameters,
such as the adiabatic constant $\Gamma$, the speed of sound $c_s$ and
others, from a paper on shocks in relativistic collisional plasmas
(Iwamoto 1989, hereafter I89, and references therein).
The speed of sound in the plasma is:
\begin{equation}
{\left({c_s\over c}\right)^2}
= \left[1+{1 \over {1+\z^2(K_2^{''}(\z)/K_2(\z) - 
(K_2^{'}(\z)/K_2(\z))^2)}}\right]
 \left[2 - \z {K_2^{'}(\z) \over K_2(\z)}\right]^{-1},
\end{equation}
where $\z=mc^2/kT$, and $K_{2}(\xi)$ is the modified
Bessel function of the second kind which naturally arises from 
the inclusion of relativistic energies in the exponent of the
partition function (I89 equation A22).

The equilibrium temperature does not quite reach its virial
value because the outwards transfer of angular
momentum, which also gives rise to radiation, is dissipative.
We therefore write the dimensionless ion temperature,
$\Theta_i = kT_i/\mu m_pc^2$, as a fraction
$\u$ of the virial temperature,
\begin{equation}
\Theta_i = \u \; \Theta_v,
\end{equation}
where $\u \sim < 1$.

The ideal gas law, which holds for collisional plasmas at high
temperatures (I89 eq. A8), is a reasonable approximation for the
equation of state, and in the ion dominated case the
speed of sound in equation (5) reduces to
\begin{equation}
c_s  \sim  \sqrt{\Gamma\u } r^{-1/2} c, 
\end{equation}
where we assumed $\Theta_i < 1$.
If the electrons in the pre-shock region are coupled with the ions,
the electron speed of sound is calculated using the exact form of
equation (5). This yields a speed which is roughly
$c_{se}\sim c/3$, because the electrons are relativistic and may be 
described by a simple equation of state as $p_e\sim \e_e /3$, where
$p_e$ and $\e_e$ are the pressure and the energy respectively. Cases
in which the pre-shock electrons are decoupled from the ions are not described
in I89 or other references, and are therefore not discussed in this paper.
We note, however, that the limit in equation (10) at the end of
this section is still valid.

An object or perturbation in the accreting plasma which is more 
compact than the
surrounding material, so that it cannot thermalize or 
lose its angular momentum on
dynamical time-scales, moves at a velocity comparable to its Keplerian 
or free fall velocity (see also section 2.3 and paper II, Sivron, Tsuruta
and Caditz 1996, hereafter STC96). The relative 
speed of such a compact object is therefore
\begin{equation}
v_* = |{\bf v}_{*} - {\bf v_g}| = 
b  \; r^{-1/2} c,
\end{equation}
where ${\bf v}_{*}$ is the velocity of the
object,
${\bf v_g}$ is the velocity of the accreting matter and $b$ is a
geometrical factor  of order unity.
The Mach number is therefor given by
\begin{equation}
{\cal M}  = { v_* \over c_s} \sim   \;  \u^{-{1\over 2}}.
\end{equation}
This Mach number is insensitive to the object's distance from the black hole
because both the speed of sound and the
typical velocity are approximately proportional to the square root of the
virial energy.
For a wide range of accretion rates $0.003<\u<1$ (I89, WL89), 
which results in
\begin{equation}
0.8<{\cal M}<16,
\end{equation}
where for the upper limit we substitute $\Gamma=4/3$. This result 
suggests that shock waves are likely to form in AGN central engines
which satisfy the constraint imposed by equation (4).
The Mach surface opening angle,
$\phi \sim 2 \arcsin (c_s / v_*)$,
may be significant (see section 4).
The shocked surface formed behind a small perturbation roughly takes a
conical shape which is likely modified by the presence of
strong gravitational and shear forces,
radiation pressure, turbulent flows and density gradients
within the central engine.

\subsection{Collisionless Shocks}

For accretion rates lower than the limit of equation (4) the deflection length
is larger than the characteristic size of the central engine,
and collision dominated shocks do not form.
However, magnetic and electric fields and turbulence are expected to be present
and collisionless shocks may occur (Kennel, Edmiston \& Hada 1985, hereafter
KEH85). In this case, the plasma state across the shock is determined not by 
the binary collisions, but by collective interactions between particles 
and self consistently
generated electric and magnetic fields. Plasma heating due to these shocks is
the result of various microinstabilities that perturb the fields and
particle distribution (Papadopoulos 1985, hereafter P85; Winske 1985).

Collisionless shocks may be present when the Larmor radius is
smaller than the central engine. The thickness of collisionless shocks is
expected to be from several Larmor radii for quasi-perpendicular shocks
to several hundred times the Larmor radius for
quasi-parallel shocks (FUR92). Note, however, that this
result has not yet been confirmed for
the near-relativistic plasma treated in this paper.
For hydrogen dominated plasmas the
ion Larmor radius is roughly
\begin{equation}
R_{Li} \sim (0.03 \,cm) {\cal B}^{-1} {\dot m}^{-1/2} M_6^{1/2}
r^{3/4} {\left({\Theta_i\over Theta_v}\right)}^{1/2}
\left({\Omega \over 4\pi}\right)^{1/2}
\left({v_{acc} \over v_{ff}}\right)^{1/2},
\end{equation}
where ${\cal B}=B/B_v$ is the magnetic field 
in units of
the virial magnetic field which can be found from 
the relation $B_v^2/8\pi=GM\rho_g/R$, and we have used equations (1) and (3).
AGN central engines may support magnetic fields of up to
the virial value of $\sim 10^4$ Gauss at $r\sim 20$ (B90).
The $\sim 10^{-8}$ gauss
fields needed for the Larmor radius to be smaller than a perturbation in the
central engine
are easily obtained from matter accreted from stars.
This limit is almost always satisfied for the fields anticipated in the
central engine (see for example B90).




For collisionless shocks in the central engines of AGN
one needs to use the kinetic equations for
relativistic plasmas.
Unfortunately such equations are not
fully developed yet (Eilek \& Hughes 1991), but several criteria
from the studies
of relativistic shocks can be used. For example, Barnes (1983) found that
Alf\'venic nonlinear waves steepen and become shocks. We therefore speculate
that superAlf\'venic motions result in collisionless shocks.


The Alf\'venic velocity is
\begin{equation}
c_A = {B \over \sqrt{4 \pi n_i m_p}}
\sim c {{\cal B} r^{-1/2}},
\end{equation}
which is comparable to $c_s$ for virial fields.
The superAlf\'venic Mach number at that distance for Keplerian speed
perturbations is thus
\begin{equation}
{\cal M}_A \sim { {\cal B}^{-1}} = B_v/B.
\end{equation}
It is therefore expected that strong collisionless
shocks may form in the central engines for moderate magnetic fields, and
moderate shocks may form for near virial fields.


Instabilities which disrupt the thin disk may also result in the
creation of a disk-corona structure in which the corona is responsible for
emitting the continuum for low enough accretion rates
(Haardt \&Maraschi 1991; Svensson \& Zdziarski 1994; ZCC95).
In the corona of a disk-corona system the corona accretion rate
is much smaller than in the tori models. In order to find the
appropriate density
for such corona one needs a model in which both disk and corona accretion are
treated consistently.
Such treatment was recently taken by ZCC95.
Using figure 5 in that paper for accretion rates
of ${\dot m}\sim 0.001$ to ${\dot m}\sim 0.1$ and using equation (12)
one finds
\begin{equation}
c_A \ge (3 \times 10^{6 - 7} cm \; sec^{-1} Gauss^{-1}) B,
\end{equation}
and thus at $r\sim 20$
\begin{equation}
{\cal M}_A \sim (3 \times 10^{3-4}) {\cal B}^{-1}.
\end{equation}

In both equations (13) and (15) the Alf\'venic Mach number generally
exceeds unity because the magnetic fields cannot quite reach their virial
value. This is because energy from the magnetic field may be utilized for 
the outwards transfer of angular momentum and/or
for dissipation, which in collisionless plasmas
may include, for example, reconnection events in the
post-shock region.

\section{Formation and Properties of the Shocked Matter}

\subsection{The Shock Creating Perturbations}

Shock creating inhomogeneities in AGN central engines may 
be the result of nonlinear growth of global perturbation in thick 
disks that tend to form planetesimals (Papaloizou \& Pringle 1984,
Narayan 1990 and references therein),
externally confined plasma sheets and filaments which cannot 
thermalize on orbital timescales 
(GR88, ST93,
and Celotti, Fabian \& Rees 1992, hereafter CFR92) or captured stars.
The simplest physical shock creating perturbation considered in this
paper is a main sequence star. A more general treatment will be 
presented in our subsequent work (STC96. See also Caditz, Sivron
\& Tsuruta 1996, hereafter CST96).
It has been shown that an unbound star can be captured 
into a tightly bound orbit around a black hole through repeated 
interactions with a thin
disk around the hole (SCR90), 
and that the probability of such an event in AGN is even
higher for stars in bound orbits 
(Pineault \& Landry 1994). Once captured the
orbit will be circularized and eventually the will be ``grounded down'' 
into the disk plane through repeated interactions with the disk.
The timescale for circularization can be shorter than that for planarization,
which may happen very close to the central source.
Therefore, in our model, the orbit of the captured star has been circularized,
whereas the orbital plane is still inclined to the disk plane.  This situation 
is expected for a wide range of
the original relative inclination angles of these planes.

The minimum conditions on perturbations with the potential to create shocks
can be determined from the requirement that they accelerate 
to supersonic velocities.  
This condition is met when the gravitational force is greater than the 
drag force at ${\cal M} = 1$:
\begin{equation}
{G M m_* \over R^2} \ge \pi D_*^2 p_*,
\end{equation}
where $p_* \sim \rho_g  {\cal M}^2 {c_s}^2$ is the pressure on the inhomogeneity surface 
(see section 3.2) and $D_*$ is the characteristic size of the inhomogeneity.
This condition can be also written in the form:
\begin{equation}
{\rho_* \over \rho_g} \ge  {R_g r \over D_*},
\end{equation}
where $\rho_*$ is the average density of the inhomogeneity. 
This condition is easily met for a typical main sequence star which is
used as an example shock producing inhomogeniety below.

The condition (17) may be even less restrictive 
if we include acceleration by radiation pressure and magnetic stresses. 
The 
amplitude and size of perturbations needed for shocks to form is therefore
rather small, and shocks may form in the wake of
large amplitude non-axisymmetrical waves in the disk. In this paper the 
possibility of nonlinear (soliton) shocks is not further 
developed because the dynamic evolution of the
perturbations is strongly dependent on dissipation (see STC96).



\subsection{Power Transferred to the Shock Front}

The power transferred from the star to the shock is at least
\begin{equation}
P_{sts} \sim  (A_* p_*) \times  v_*,
\end{equation}
where $A_*=\pi D_*^2$ is the cross section area of the star and $p_*$ is the 
pressure near the stellar surface, behind the shock.

The pressure at the leading edge of the stellar surface can be estimated
from the treatment of shock
formation  in supersonic non-relativistic flows past solids of
revolution (Landau \& Lifshitz 1987, hereafter LL87):
\begin{equation}
p_* \geq 
1.5 \rho v_*^2 {1 \over [1 - 1/(5 {\cal M}^2)]^{1.5}}.
\end{equation}
The adiabatic constant was chosen to be
its highest value of $\Gamma = 5/3$, which results in a lower limit on $p_*$. 
The power transferred from the star to the shock is thus at least
\begin{equation}
P_{sts}  \sim  (1.5 \times 10^{43} ergs \; sec^{-1}) \,
{\dot m M_6\over r^3}
\left({4\pi \over \Omega}\right)
\left({v_{ff} \over v_{acc}}\right)
\left({D_* \over R_g}\right)^2.
\end{equation}
The radiation thus emitted can reach the observed luminosity with 
a {\it{single large perturbation}}, if the 
shocked material radiates efficiently. 
For collisionless shocks the ambient pressure is modified to be
\begin{equation}
p_A= p_*+ {B^2 \over 8\pi}, 
\end{equation}
and the pressure from equation (19) is somewhat larger, especially 
with $\Gamma\sim 4/3$
for the heat capacity ratio when the magnetic effect is dominant.

We next argue that if the accretion is already near its Eddington limit
shocks will provide
the mechanism for making the post-shock region radiative. We separately treat
the collisional and collisionless shock cases.


\subsection{Radiative Efficiency of Shocked Material}

\subsubsection{Collisional Shocks}

In the near-Eddington regime radiative processes in the post-shock region 
may be extremely efficient, due to, for example, 
bremsstrahlung from thermal pairs. 
We speculate that most of the energy is radiated by
pairs which are created in the post-shock region. 
The presence of some pairs is expected in collisional hydrogen
plasmas with bulk motion with velocity of order
\begin{equation}
U\ge\left[{2m_e c^2 \over m_p}\right]^{1/2},
\end{equation}
(see I89) and in the central engine the bulk motion is of order $v_*$, 
which is generally much larger than $U$ in equation (22).

A more significant pair-creation is present in the 
post-shock region because 
the temperature of the post-shock electrons is $\Theta_e^{'} >> 1$,
and because these electrons are thermalized.  We hereafter 
denote with a prime the post-shock quantities.
In this scenario the optical depth for 
pair-annihilation is larger than unity, and a high density pair equilibrium is
established in the post-shock region. These circumstances were investigated by
Done \& Fabian (1989, hereafter DF89), CB90 and others. In their simulations these authors found that the 
spectrum, which is a power-law due to, for example, Comptonization of 
bremsstrahlung radiation, is flattened because of
the increased compactness of the source.

This situation is also likely to apply locally to
the post-shock region because the optical depth for 
pair annihilation is of order 
$\tau_{pa}\sim 0.4 n_l^{'} \sigma_T \lambda = 5 \Theta_e^2$
and  $\Theta_e \ge 1$.
The timescale for mildly relativistic ($\Theta_e\sim 1$)
electrons to thermalize
$t_{e}\sim 1/(c \sigma_T n_i^{'} \ln \Lambda)$ is
shorter than the shock crossing timescale 
$t_{sw}\sim \lambda/c_s$, and, using eq (2), we find the following ratio 
to be smaller than unity 
\begin{equation} 
{\left[{t_e \over t_{sw}}\right]} = {3 \over 64} {c_s\over c} {1\over \Theta_e^2} \le 1.
\end{equation}
Here we have used equation (5)
and the condition 
\begin{equation}
{n_{i}^{'}\over n_{i}} \ge 4,
\end{equation}
from I89.  The condition (23) is generally fulfilled, 
and the post-shock electrons are therefore thermalized
(see also CST96, STC96, for a more detailed approach).

For the thermalized post-shock pair equilibrium described above,  
the total number density of pairs, 
$n_l^{'}$, may be estimated by evaluating the post-shock
energy of the ions and assuming equipartition. The ion energy is approximatley
\begin{equation}
{\epsilon_{i}^{'}\over m_e c^2} \ge 4 n_{i} 
\left({1 + {m_p\over m_e}{3 \Theta_{i}^{'}\over 2}}\right),
\end{equation}
where we have used the post-shock energy from I89 and equation (24).

At $r\sim 20$ the bulk velocity is roughly $v_*\sim c/\sqrt{20}$, and
the post-shock temperature is roughly 
$\Theta_{i}^{'}=30 m_e/m_p$, which translates to more than 
$\sim 40$ electrons and positrons per post-shock ion in
equation (25), and we have used equations (8) and (22). Our assumption 
that at the above location the post-shock creation of pairs is 
unstable for ${\dot m}\sim 0.1$ is supported by Bjornsson \& Svennson 1991
(see also figure 1).

Because the density of pairs in the post-shock region
is so much larger than that of ions in the 
pre-shock region, and because bremsstrahlung radiative emmisivity is 
proportional to the square of the density,
the post-shock pairs may lose most of their energy to bremsstrahlung radiation.
The total bremsstrahlung power per unit volume is of order
\begin{equation}
j \ge (1.2 \times 10^{-23})erg \; cm^{-3} sec^{-1} {\left({n_i}^{'}\right)}^{2} 
{\left({\Theta_e}^{'}\right)}^{1/2},
\end{equation}
(e.g. FKR92).
The volume of emitting post-shock matter is of order $\lambda \pi D_*^2$,
and the power radiated assuming an optically thin source is 
\begin{equation}
P_b \sim j \lambda \pi D_*^2,
\end{equation}
which, using equations (20), (25) and (26), generally exceeds $P_{sts}$.
The power in equation (20) is therefore emitted as bremsstrahlung radiation if
$P_b\ge P_{sts}$, which is equivalent to the condition
\begin{equation}
(\Theta_e^{'})^{1/2} {\Theta_e^2} \ge  0.1 r^{-1},
\end{equation}
which is usually satisfied, and we have used equations (2), (3) and (20).
In figure 1 we show that the increase in temperature and 
electron--positron number density across 
a typical central-engine collisional shock results in an increase by
several orders of magnitude in 
bremsstrahlung emission rate across shocks of moderate to high Mach numbers. 
The Ranking-Hugoniot relations are taken from I89, the density of pairs 
in both the pre-shock and post-shock regions
is taken from Svensson 1982, Svensson 1984 and 
Bjornsson \& Svensson 1991, and the 
typical parameters of the two-temperature plasma are taken from WL89 
at a distance of $r\sim20$.
This figure is taken from a more general discussion in CST96. Figure 1 
gives the lower limit on the post-shock radiation, because additional
increase in emission due to, for example, Comptonization of the 
bremsstrahlung radiation, 
for example, is not taken into account.
We therefore conclude that the post-shock region can easily radiate away most
of the power transferred from the star to the shock. 
We use this result in section 4 for deriving a possible light-curve 
from an emitting shock.

It has been suggested that for certain ranges of compactness the 
hot gas in the central engine is unstable, and the perturbations
may result in the formation of a `two-phase'  region consisting of a dense
cool (not completely ionized) phase and a hot phase (GR88). 
In a similar manner the post-shock
matter in our model may also form two phase configuration.
The post-shock compactness is of order 
\begin{eqnarray}
{\epsilon_i^{'} \over m_e c^2 n_i^{'}} &\sim & {P_{sts} \over \lambda 
2 \pi D_* c_s n_i^{'} m_e c^2} \nonumber \\ && \\  
&\sim& 
1600 {{\dot m}\over r^{5/2} {\sqrt{\Gamma \u}}} \Theta_e^{-2} 
\left({D_* \over R_g}\right) 
\left({\Omega \over 4\pi}\right)^{-1}
\left({v_{acc}\over v_{ff}}\right)^{-1}, \nonumber 
\end{eqnarray}
where we have used equations (3) and (20), and assumed $\u \sim 0.01$.
This compactness is related to the standard compactness, $l = L \sigma_T
/(R m_e c^3)$ (Guilbert, Fabian \& Rees 1983).
As in GR88, the high compactness leads to an interesting effect: 
In some post-shock regions ions are trapped and may 
lose most of their energy to the surrounding electron--positron pairs through 
a runaway cooling which involves 
Comptonized bremsstrahlung radiation and then bound-free transitions. 
Counter to our experience with other shocks 
some post-shock ions after being initially heated may cool below the 
original pre-shock temperature to a typical 
GR88 cold phase temperature of 
\begin{equation}
 T_{ps} \ge \left(\left({1-A}\right){L\over 
{16 \s \pi R^2}}
\right)^{1/4},
\end{equation}
where $T_{ps}$ is the post-shock temperature and 
$A$ is the flux averaged albedo.
This matter, which probably forms clouds of denser cold matter, may be `warm', 
i.e. partially ionized, because of its close proximity to 
the strong radiation from the nearby shock, or `cold' if the lifetime
of the cold matter in the cloud is longer than the Keplerian timescale
and the matter is not intenesly irradiated.


A configuration with cold plasma in
the central engine may therefore exist. The cold component may
give rise to some emission and
absorption features such as the gravitationally red-shifted
Fe K line recently observed by ASCA (Tanaka et al. 1995). 
The cold (or warm) phase may also eclipse the hot plasma under 
certain conditions. The eclipse may 
modify the primary emission by absorption in a post-shock foamy
sheet behind the Mach surface if
$P_{sts}$ is expended on the cooling of the post-shock matter,
and if the lifetime of the cold blobs is long 
enough (see appendix).  In subsequent papers (CST96, STC96)
we will present a full numerical solution for
the rate equations in the two-phase
post-shock region which accounts for these effects.


\subsubsection{Collisionless Shocks}

A naive application of the same procedure used in 3.3.1 to collisionless
shocks would have us compare the shock crossing time, which is either 
$t_{sw}\sim r_{Li}/c_A$ or the possibly longer 
$t_{sw} \sim (r_{Li} / c_A)(1+ \omega_b r_{Li}/c_A)$ which is 
increased due to confinement (where $\omega_b=eB/mc$
is the cyclotron frequency), to the deflection time 
$\lambda /{\sqrt{2KT/m}}$ (which is the typical timescale
for, for example, conduction losses).  Sharofsky, Johnson 
\& Bachynski 1966 used these timescales $t_{sw}$
for the diffusion time out from a confined region.
The problem with this approach is twofold: 
(i) The width of a collisionless shock tends to 
be at least several Larmour radii (e.g. KEH85) and the crossing
time is therefore several times $t_B$.
(ii) As discussed in section 2.2 the plasma losses across collisionless
shocks are determined not by single binary collisions, 
but rather by collective 
interaction between particles and self consistently generated electric 
and magnetic fields.
Most of the losses in collisionless shocks, (especially in the near
virial fields limit), may be due to, for example, various micro-instabilities 
in the collisionless plasma rather than conduction (Winske 1985, 
P85). 

At present there are no analytical Ranking-Hugoniot relations for relativistic
collisionless shocks in matter with density in excess of 
$n\sim 10^{10}$ cm$^{-3}$. We therefore parametrize the increase in 
magnetic field magnitude, and possibly particle density and velocity, 
across the shock. This increase would result in an increase of the 
synchrotron and synchrotron-self Compton radiation, 
and would take place over a timescale similar to $t_B$.
 
We speculate that the Ranking-Hugoniot relations 
for the conditions envisioned in the central engine have some common features
with the Ranking-Hugoniot relations of lower energy collisionless shocks 
(e.g. P85 and reference therein). 
For simplicity sake we assume that the quiecent radiation is 
due to synchrotron radiation only, 
such that 
\begin{equation}
L(0)\sim 2 \times 10^{-9} B^2 \gamma n V_X,
\end{equation}
as in B90, where $L(0)$
is the actual observed bolumetric luminosity and
$V_X$ is the volume of non-thermal emitting medium, which can be 
either in a quasi-spherical form or in a corona-over-a-disk 
form (see section 2).
The resulting post-shock radiation density is therefore of order
\begin{equation}
P_{s}\sim \left({n^{'}\over n}\right) \left({\gamma^{'} \over \gamma}\right)
\left({B^{'}\over B}\right)^2
{L(0)\over V_X}.
\end{equation}
We hence estimate the overall compactness of the post-shock matter 
for quasi-spherical accretion:
\begin{equation}
{P_{s} \sigma_T R_X^2 \over m_e c^3} = 10  
M_6^{-1}\left({{\dot m}\over 0.01}\right) \left({r\over 20}\right)^{-1}
\left({n^{'}\over n}\right) \left({\gamma^{'} \over \gamma}\right)
\left({B^{'}\over B}\right)^2,
\end{equation}
and for coronal accretion:
\begin{equation}
{P_{s} \sigma_T R_X h \over m_e c^3} = 10 
M_6^{-1}\left({{\dot m} \over 0.01}\right)\left({h\over 20 R_g}\right)
\left({r\over 20}\right)^{-1}
\left({n^{'}\over n}\right) \left({\gamma^{'} \over \gamma}\right)
\left({B^{'}\over B}\right)^2.
\end{equation}
Extrapolating from interplanetary collisionless shocks (KEH85) 
and extragalactic 
collisionless shocks in jets (see Eilek \& Hughes 1991 and references therein)
we speculate
that the post-shock magnetic field $B^{'}=\kappa B$ and density $n^{'}$
are larger than the corresponding pre-shock parameters.
The compactness may therefore be increased enough for pair production
to be important. The resultant pair
production will be non-thermal, and the steepening of the spectrum of 
emission from the post-shock region may be significant. This post-shock 
emission will result in steepening of the overall spectrum if the post-shock 
emission is comparable to that of the quiscent emission.
Given a large perturbation and a strong shock with high ${\cal M}_A$ number
(see section 2.2) the post-shock parameters may result in a significant 
increase of the flux and the steepening of the spectrum.
We discuss the implication of this result in STC96.

\section{Modification of the Continuum Source}

\subsection{Effects of Radiative Shocks}
Since the speed of the
`star' in our model is assumed to be roughly $\sim 0.2 c$ at $r=20$ we will
neglect the increase in shock radiation due to the Doppler effect (for 
corrections see STC96). We note,
however, that the angle spanned by the shock front area vector and the
line of sight determines the percentage of outcoming shock radiation due
to the post-shock eclipsing which is dealt with in the next subsection.
Therefore emission only doubles when the shock front is
directed towards the observer (see Figure 2).
In our approximation we assume that the spectral index in the post-shock 
pair-dominated region is $\sim 1$, 
as was shown by DF89, CB91, TK95 and
others, whereas in the ambient matter the 
spectral index is $\sim 0.5$, as is expected in synchrotron and 
synchrotron-self Compton sources (B90 and references therein). 

In Figure 3 we show the effect of a shock on the
spectral index in the soft X-ray wave-band in the $0.2$--$2$
keV window,
the hard X-ray wave-band in the $10.0$--$20.0$
keV window and the very hard $100.0-300.0$ keV window. 
The angle $\b$ is spanned 
by the shocking star, the central engine and the observer-black hole line of
sight.
One can clearly see that the outcoming flare is soft 
as observed in several compact seyfert galaxies. The lesser 
softenning in the higher energy bands is also consistent with observations
(see MDP93).
We assumed that the maximum observed flux with spectral index of order 
unity in the $0.1$--$300$keV range is similar to the flux from the 
abient matter.
We note that a spectral break near $300$keV was assumed in our work because 
we assumed a 2D slab-corona configuration for our pair spectrum,
as in TK95, which results in a spectral break near
$\sim 300$keV for high enough compactness. A combination of observations
with present and future missions, such as XTE, XMM and AXAF, 
may be able to determine if the light-curves fit these characteristics.

\subsection{Effects of Post-Shock Cold Matter}
 
Here we obtain the light-curve for radiation from
hot plasmas in the central engine which is `eclipsed' by the cold 
post-shock matter. The central engine will be eclipsed when the post-shock
cold matter creation timescale, of order $t_e$ in equation (23), 
is shorter than the cold matter timescale $t_s$ (equation A9 in appendix), 
and when the post-shock matter column density is of order 
$n^{'} c_s t_{s}\sim 10^{21-24}$cm$^{-2}$  (depending on the 
ionization state of the post-shock `cold' matter). 
Both contitions are met for post-shock matter in at least some 
AGN (see equations (2), (3), (7), (24) and (A9)), 
and an eclipse of the central engine is therefore possible.

In obtaining the light-curve our approach is geometrical in nature.
We assume that a conical shock front, similar in shape to a simple
Mach cone (see section 2) eclipses a circular continuum source.
To highlite the geometrical effects we also chose a rectangular continuum
source. A possible picture of what happens near the central region
of an AGN with an inhomogeneity is shown
in Figure 2. Here a shock inducing star 
(at the tip of the cone) orbits inside a
thick accretion torus which is located outside the X-ray continuum region, but
confined to a region smaller than the thin outer disk 
(points F1 and F2 are the
inner edges of the thin disk.) From
an observer's point of view the points O, P and Q on the Mach cone (Figure 4)
form a `fan' in the shape of a triangle that partially obscures the 
internal continuum source.
The type of occultation depends on the optical depth of the
matter on and near the shocked surface.

Since the extent of the optically thick 
shock may be limited, as was shown in section 3, we choose the
point of view of an 
observer in direction G, which requires
the shortest extent of optically thick matter.
Figure 4 is a simplification of the observer's view 
which is used for the calculation of 
the light curve. It is based on the
following assumptions.

\noindent
1. The continuum X-ray emitting region is assumed to be circular,
with homogeneous surface brightness, as viewed when unocculted. 
Its radius is $R=R_X$. We note that in an
alternative model, in which the soft X-ray region is larger than the
hard X-ray region, the soft minimum 
flux will be significantly delayed with respect to the hard flux.

\noindent
2. The Mach surface is assumed to be an equilateral triangle, passing 
across the
center of the circular source.

\noindent
3. Density variations are simplified by assuming three density
regions. In region 1 (angles  smaller than $\phi_1$) 
the matter is opaque to both hard and soft
X-rays. In region 2, with angles between 
$\phi_1$ and $\phi$,
the matter is transparent to hard X-rays, but opaque
to soft X-rays. In region 3 the density of matter varies with distance,
due to the spread of matter, and is partially transparent. 
We consider only the hard X-ray wave-band in the 10--20
keV window in which no absorption is expected
and the soft X-ray wave-band in the 0.5--2
keV window in which absorption due to cold or warm oxygen is expected. 
We chose the maximum covering factor to be $f=0.5$ for 
demonstrative purposes.

\noindent
4. The strength of the pre-dip flare was assumed to be negligible for 
demonstrative purposes. 

In Figure 5 we see the light curves near a dip in two energy bands
that will be observed when the
source is eclipsed by the post-shock matter. With a Mach surface 
opening angle of $\phi=45^0$ the softer X-rays light curve
is dash-dotted and the harder X-rays light curve is solid. 
The dotted curve is for the case of a smaller
opening angle of $\phi=30^0$, for the hard X-rays. The dashed curve
is for the simplified case in which a triangle with an opening 
angle $\phi=45^0$ eclipses a square source. In all cases
the maximum covering factor is $f=0.5$.

The decrease of flux toward the dip is energy dependent. A delay is observed 
in the sense that the decrease in photon count in higher energies lags 
the decrease in the lower energies. The light curve in the
hardest window has a cusp--like shape, 
dropping to half its
value at the bottom. The bottom was chosen to be at half the intensity
so that the size of the central sector of 
absorbing material within $\phi_1$ which is opaque to the harder X-rays is 
approximately half the size of the harder X-ray source size 
during maximal occultation. 
The maximum value for $R_X$ is thus obtained with $\phi_1= 45^0$. 

\section{Discussion}

The results of sections 2, 3 and 4 imply that given a
perturbation, an accretion rate and a black hole mass one could 
calculate temporal features and confirm the `condensation'
of cold (not completely ionized) matter.
Our work thus complements the work by GR88 and LW88. 
We also argue that one could also calculate the variability and 
power spectrum, the rate of angular momentum transfer and the 
effects of particle acceleration. 
The results of sections 2 and 3 may give rise to the prediction of the
$r_{hs}=l_h/l_s$ ratio, $l_h$ and $l_s$ 
where $l_h$ is the compactness of the hard 
non-thermal radiation
and $l_s$ is the compactness of the soft thermal `seed photons' (see STC96).

When the accretion rate is relatively 
low one expects a disk-corona configuration. In this configuration accretion
disk instabilities are expected to yield microvariability (e.g. W93). 
As argued in sections 2 and 3 collisionless shocks will then
tend to spread these perturbations throughout the corona at the fastest
magnetosonic speed, thus enabling large amplitude variability. 
Since the coronal compactness is relatively small only a fraction
of the perturbation energy will be radiated away (and most of the perturbation
may travel outwards as winds). Note, however, that various microinstabilities
may increase the efficiency of energy transfer.

The radiation will tend to be relatively hard 
since no Comptonized bremsstrahlung radiation is expected and only
synchrotron-self Compton (SSC) or non-thermal pairs cascades are 
possible locally, 
with a very hard outcoming spectrum 
(e.g. B90, TK95). 
The compactness $l_s$ also depends on 
${\dot m}$ (FKR92), and $r_{hs}$ may not change 
as fast as $l_h$. The radiation may be SSC only if $l_h+l_s$ is such 
that small IR blocking cold cloudlets form in the shocked material 
(as in CFR92), or else IR-X-rays correlation should 
be observable (Done et al. 1991). Reflection of the hard radiation from the
relatively cold disk component may result in an Fe emission line, such as the
one recently observed by ASCA (Tanaka et al. 1995 and references therein).

When the accretion rate is relatively high (${\dot m}\geq 1$)
one may expect a very dense corona, or a thick two-temperature torus. 
Perturbations will be the result of instabilities in the disk (e.g. Narayan
1990 and references therein) 
and the resultant shocks will produce bremsstrahlung-Compton
radiation or may even result in a two-component Guilbert \& Rees medium. 
The emission 
from such media becomes softer when $l_h+l_s$ become higher
(ST93, Nandra \& George 1994). In general one expects 
$l_s$ to increase more than $l_h$ in this case, because of the soft photons
that originate or are reprocessed in the cold component.

The emission spectrum for the high ${\dot m}$
is dominated by a more thermal-like 
component with a possible high temperature bremsstrahlung, which may imply 
AGN evolution which is thought to be responsible for the 
observed x-ray background.

Shocks can naturally explain a common feature of the spectra
of most AGN.
In the spectra of radio quiet AGN the ratio of thermal to 
non-thermal emission is probably of the order 
$0.3 \le {L_T \over L_{NT}} \le 3$,
where $L_T$ is the thermal luminosity and $L_{NT}$ is the power-law
non-thermal luminosity (LW88). This result is
readily explained if the hot continuum emitting plasma is 
nearly optically thick (Sivron 1995).
A mechanism in which the optical depth is {\it{limited}} to 
unity involves a two phase medium: A medium that would have been
optically thick if the accretion were spherical becomes a two phase medium 
with decreased optical depth for the hot phase whenever the optical thickness
exceeds unity, or equivalently whenever the luminosity is near-Eddington 
(ST93, STC96, GR88). 
Such two phase media are naturally produced 
with shocks, because shocks are so readily available in the central engine
(as was shown in section 2), and because shocks naturally channel the energy
of perturbations into radiative cooling which cools portions of the hot phase
in the case of near-Eddington luminosities (as was shown in section 3). 
The cycle is completed when the speed of sound in the rarefied 
hot phase again falls below the Keplerian velocity due to the `evaporation'
of cold phase clouds,
which is usually faster than the dynamical
timescales, or due to cooling processes in 
the hot plasma, and shocks are again possible 
(for a typical timescale of `evaporation' see, for example, 
Zeldovich \& Raizer 1967 pg 571). 

If radio quiet AGN central engines have a disk-corona geometry, 
the observed spectra cannot be explained by homogeneous coronas 
(Fabian et al 1995, Stern et al 1995, Iwasawa et al. 1995). 
This paper establishes the presence of shocks that 
naturally distrupt the homogeneous corona. 
Perturbations in the hot homogeneous coronas may be considered in 
the sense described in this paper in order to
establish the true geometrical shape of the 
inhomogeneous corona and calculate the resultant spectrum.

Shocks may also explain the observed power spectrum of the central engine.
A weakness of our assumption that a single star is responsible for the shock
and resultant emission is that periodicity, which has not been
observed in AGN, may be expected. 
A star orbiting a supermassive black hole can be a curiosity, or a
rule --- depending on future observations, but
from section 2.3 it is apparent that 
any large perturbation present in the 
innermost $20 R_g$ can yield the luminosity change
of equation (20). It is therefore 
probably more likely that the power spectrum is the result of 
perturbations which are not associated with a captured star.
If the conditions in equations (4) and (30) are met for these perturbations,
and if perturbations of characteristic size, time
and brightness occur
at a random distance from the black hole,
the resultant power spectrum slope
will be between that of random shot
noise and chaotic noise.
Our subsequent work (STC96) uses the arguments of
Sunyaev \& Titarchuk (1980) and DF89 to obtain these
power spectra which are comparable to 
the power spectra observed in radio quiet AGN (Green 1993 and references
therein). 
The condition in equation (30) on the compactness then determines 
that a one dimensional introduction of 
perturbations, rather than the two dimensional 
introduction of perturbations as in thin disk models, 
is needed in order to reproduce 
the observed power spectra. The physics of random introduction of 
perturbations in one
dimension may be easier to explain than the random introduction 
of perturbations in two dimensions (STC96).
Condition (29), or an equivalent for a non-spherical perturbation, may result
in a high frequency cut-off in the power spectrum (STC96).

If the shocks are collisionless the source may not be optically thick, 
and modification through pair-creation may not be as significant, unless
the fields are near virial.
In the case of collisionless shocks that do not satisfy the conditions 
in equations (32) -- (34), one should note that energy may still be deposited 
by the shock to the ambient hot plasma. Various microinstabilities may 
still result in micro-variability, but large amplitude variability is 
harder to produce. 

In the above discussion of variability 
we did not try to calculate the timescale in which 
the post-shock matter returns to its original state. We 
generally assumed that the recovery timescale is shorter than the dynamical
timescales. If the recovery timescale is larger than the dynamical 
timescale the power spectrum is further modified. 
The effect of a two phase medium on the variability is also calculated
in our subsequent work (STC96).

The spectral variability of flares and dip
described in sections 3 and 4
was already found to be similar to that of several Seyfert 1 galaxies 
such as NGC 4051 (Kunieda et al. 1990) and MCG -6-30-15 
(Nandra et al. 1990, Matsuoka et al. 1990, Fabian et al. 1994, Otani et al.
1995, 
Sivron et al. 1996) and others (MDP93).
The similarity is not only in the sense that the power-law spectral
slope is anti-correlated with the overall flux, but also in the 
sense that the low flux level is correlated with some 
emission and absorption features. Although this correlation was traditionally
explained to be the result
of matter in the line of sight which is far from the central engine (e.g.
Nandra et al. 1990) it is generally hard to explain why 
this matter is always smaller in 
thickness than the size of the central engine (Sivron et al. 1996).
The emission and absorption features may thus be associated with an 
increase in optical depth of cold matter within, or very close to, 
the central engine, as in the case described in section 4.2.

Shocks and their byproducts may be used for the outwards
transfer of angular momentum in the central engine 
(STC96). 
Spiral shocks, which were considered for the outward transfer of angular
momentum in the outer disk, may form when a massive companion perturbs 
the disk (Savonije, Papaloizou and Lin 1994, and references therein, 
in reference to  galactic sources). Angular momentum can be transferred in 
this way in AGN inner disks as well (Chakarbarti \& Wiita 1995).
Spiral shocks probably take a completely 
different geometrical shape in compact central engines, 
where the geometry of the emitting 
material is not that of a thin disk. 
The angular momentum transfer in the central 
engine in high accretion Seyfert 1
galaxies and radio quiet quasars can be due to chaotic, rather than 
spiral, shock structure (STC96).

In the case of lower accretion rates significant outflows must also be 
considered as a part of the total
angular momentum transfer, since the optical depth is limited and mass ejection
may become significant. However, the
availability of shocks in the central engine and the presence of mechanisms
that can disturb the axisymmetrical structures 
makes it imperative that shock related angular momentum transfer is
significant. Examples of mechanisms that can disturb the axisymmetry 
include a poloidal magnetic field structure, 
a Lense-Thirring precession of tilted axisymmetrical structures 
in a Kerr geometry, etc.

Shocks may also be an indirect agent in the outward transfer of 
angular momentum. The post-shock cold clouds may transfer angular momentum
through binary and triple collisions (STC96). Because the 
hot component in a GR88 two phase model is fairly optically thin, some 
cold clouds may also be ejected carrying with them large amounts of 
angular momentum. 
 
Other processes 
affected by the presence of shocks include particle acceleration
and pair production. Many models use an
input energy spectrum for a large non-thermal 
electron population in order to obtain their
model spectra (B90 and references therein, 
CB90). 
From our shock model one should directly calculate the 
normalization of this distribution by utilizing, for example, the theory of
Fermi and shock-Fermi acceleration for a specific post-shock pre-shock
compression ratio (Blandford \& Eichler 1987, Levinson 1994).
We note that modifications due to Fermi acceleration
will probably result in the lowering of  
cooling efficiency, as in equation (30). 
That a significant portion of the particles may gain a non-thermal high energy
distribution function was anticipated (but not 
calculated) by many previous authors that discussed pair modifications
of spectrum and structure of the central engine 
(see Guilbert, Fabian \& Rees 1983, Begelman, Sikora \& Rees 1987, 
Tritz \& Tsuruta 1989, Kusunose \& Takahara 1989, CB90,
GH94). 

\section{Conclusions}
We have shown that shocks in the central engines of AGN form for a wide range
of accretion parameters. We have also shown that perturbations in the central
engine will effectivly radiate their energy away through shocks, and that
for a wide range of accretion parameters that radiation will be observable. 
Given the generality of our results we believe that shocks should be 
fairly common and that their effects on observations should be significant.

Acknowledgments:

We thank the anonymous referee for some very helpful remarks. We thank 
Dr. M.J. Rees, Dr. N. Iwamoto and Mrs. N.Sivron for useful 
discussions and comments. We also thank Mr. W.McHargue and Mr. M.J.Kellen for 
their help with supporting software.
This work was supported in part by NASA grants NAGW-2208 and
NAG8-230 and NSF grant RII-8921978.

\appendix
\section{Appendix}
The lifetime of the thermal gas sheets and 
filaments
that form behind the shock front can be evaluated using one of the two
following methods:

The minimum lifetime of plasma sheets and filaments can easily be 
approximated for the 
case in 
which the `sheets' are in the shape of a spherical cloud 
and all of the incident radiation is 
transformed into an energy increase of size $\Delta E$:
\begin{equation}
t_{*} \ge {\Delta E \over F} =
140 r^{1/2} \; \left({r_c \over c}\right) \; 
\left({k T_c \over \m m_H c^2}\right),
\end{equation}
where $ F$ is the maximum flux incident on a cloud, $r_c$ is the radius of a 
cloud and $T_c\sim 10^8$K is the 
final temperature of the clouds in which all the elements are completely
ionized (see also Bond \& Matsuoka
1993). For a typical cloud of  
size $10^8$cm (for $M\sim 10 M_6$, using the same arguments as in 
CFR92),
the lifetime of the clouds is greater than 
$10^{-4}$ seconds. This result is a lower limit, because no radiative 
losses are 
considered in spite of the fact that bound-free and free-free emissions,
which are the dominant mechanisms
in the creation of such thermal matter,
have already been
proven to be efficiently radiating energy in section 3.2. Whereas this 
timescale describes the dissipation of a cloud, we next try to 
determine the beginning of such a process, which we show to have 
longer time-scale.

We assume
that the temperature decrease is due to 
adiabatic expansion into vacuum,
and calculate the subsequent timescale necessary for the optical 
depth to be halved.
The following equations govern the adiabatic expansion of clouds.
The continuity equation in spherical coordinates, for a spherically symmetric 
time-dependent case (LL87):
\begin{equation}
{\partial \rho \over \partial t} + {1 \over r^2} {d \over dr} (r^2
\rho v) = 0,
\end{equation} 
where
r is the radius of the exploding plasma cloud and v is the velocity of a fluid
element. The Euler equation
\begin{equation}
{dv \over dt} + {1 \over \rho}{dP \over dr} = 0.
\end{equation}
The adiabatic (energy) equation is
\begin{equation}
P=T^{\g / (\g - 1)},
\end{equation}
where $\g=5/3$ is roughly the adiabatic index for an ideal 
monoatomic gas, for Hydrogen dominated plasmas.
The equation of state is:
\begin{equation}
T \sim {m_H P \over \rho k_B}.
\end{equation} 
The total mass of a cloud, $M_{cld}$, is
\begin{equation}
{M_{cld}}={4 \pi \over 3} r^3 \rho.
\end{equation}
The optical depth of a cloud, assumed here to be bremsstrahlung 
and bound-free dominated, is:
\begin{equation}
\tau \sim  1.0\times 10^{3} r T^{-1/2} Z^2 \rho^2 \times \left({h \nu 
\over KeV}\right)^{-3}.
\end{equation} 

The expansion of the clouds results in their transparency. The
optical depth of the clouds as they expand (and are assumed to be 
opaque, $\tau=1$, at that size), can be related to the corresponding 
variations in cloud size, density and temperature without the need 
to solve equations (A2) and (A3) explicitly.
By substituting equation (A6) into (A5) and than into (A4), we get
$T\propto r^{3-3\g}$, and substituting the result into (A7) we find
\begin{equation}
{\tau (r_0) \over \tau (r_f)} = \left({r_f \over r_0} \right)^6.
\end{equation}
Here $r_f$ is the final size of the expanding cloud.
From this we can see that a bremsstrahlung and bound-free dominated, 
adiabatically expanding
cloud will only need to grow by $r_f/r_0 \sim 1.12$ for the 
optical depth to decrease by a factor $2$. Substituting these initial and 
final sizes into equation (A3) we can 
estimate the lifetime of a cloud. Since the pressure, density and size
change very little
when a cloud opacity is halved the Euler equation
becomes roughly
$ d^2r / dt^2 \sim - (1/\rho_0) (P_0/r_0)$. Integrating
twice, using the initial conditions $v(t=0)=0$ and $r(t=0)=r_0$, we get 
\begin{equation}
t_s > \sim (100 sec)\times {(r_8)^2 \over T_6},
\end{equation}
where $r_8=r_0/10^8$cm, $T_6=T/10^6$K, $r_0$ and $P_0$ are the initial 
size and pressure of the dense sheets, and $t_s$ is the time from 
the beginning of expansion.  Hence, the post
shock sheets become transparent
on a timescale of hundreds of seconds. As a result a `curtain' of cold
matter of thickness $t_s c_s$ is comoving with the mach cone. This curtain
may partially obscure the radiation from the central engine if its optical
depth, $n' t_s c_s \sigma$, where $n'$ is the post shock cold matter
average density and $\sigma$ is the cold matter's effective cross section,
is nearly unity.
Because of confining pressure from
magnetic fields which may appear 
on the right hand side of
equation (A3) the lifetime of a cloud may be longer than that in equation (A9).

Considering the velocity of the inhomogeneity
found in section 2 and the fact that the post shock matter moves at 
subsonic velocity, this result determines that the 
clouds form a `foamy' wall
of thickness $\sim c_s t_s \sim 10^{10}$ cm. A typical filling factor of
$C \sim 0.01$ is reasonable for the ratio of ambient and thermal
matter under the assumption that half the ambient matter becomes
thermal in the post shock region (with the necessary covering factor
$\sim 1$, see CFR92, ST93).
Therefore, on the average, one sheet of thickness
$10^8$ cm can partially eclipse the continuum source.

\clearpage
%

\begin{figure}
\caption{The effects of the increase in temperature, 
number density and number of pairs across a typical central-engine 
collisional shock with Mach number $M$ results in an increase in 
bremsstrahlung emission rate. 
The post-shock to
pre-shock ion temperature, ion density and bremsstrahlung emissivity 
ratios, $T'/T$, $N'/N$ and $b'/b$,
are the dashed, dotted dashed and solid curves respectively. 
They are plotted versus the Mach number.
To the right of the thick solid line is the pair-runaway region where 
condition (25) is fully satisfied. 
The pair rate equations the pre-shock
and post-shock regions in our code i
are taken from Svensson 1982, Svensson 1984 and 
Bjornsson \& Svensson 1991. The Ranking-Hugoniot relations for relativistic
collisional plasmas are taken from Iwamoto 1989.
This figure is taken from a more general discussion in 
CST96. The pre-shock temperatures are taken from WL89 for 
${\dot m}\sim 0.1$ at $r\sim20$. The post-shock ion temperature
to electron temperature ratio of roughly $\sim 10$, is
appropriate for $\log N'/N \sim 0.6$.}
\end{figure}

\begin{figure}
\caption{The doughnut shape ion-torus shown here (not a linear scale) 
consists of an inner part of a thin disk (at points F1 and F2), the central 
black hole (at point C) and the continuum X-ray source (shaded in gray,
near the black hole). The rest of the ion-torus is filled with optically thin
two-temperature plasma. Inside the torus a star orbits, and shocks the
two-temperature plasma.  When the shock front moves towards the observer a 
flare is seen. Behind the surface of the `Mach-cone' optically
thick matter forms, which may obscure the continuum region when in the line
of sight. The obscuring is most effective if the observer is in the
direction of point G.} 
\end{figure}

\begin{figure}
\caption{The effect of a shock on the spectral index of flux from the 
central engine with a radiative shock front is plotted versus the angle 
spanned by the Keplerian perturbation, the central black hole and the 
line of sight. The solid line is for flux in the `soft' $0.2$--$2$keV window, 
the dashed line is for flux in the `hard' $10$--$20$keV window, and the 
dotted-dashed line is for the $100$--$300$keV window.
The trend of harder flares in the higher energy bands is apparent. At an angle
$\pi /2$ the Mach cone is pointing towards the observer.}
\end{figure}

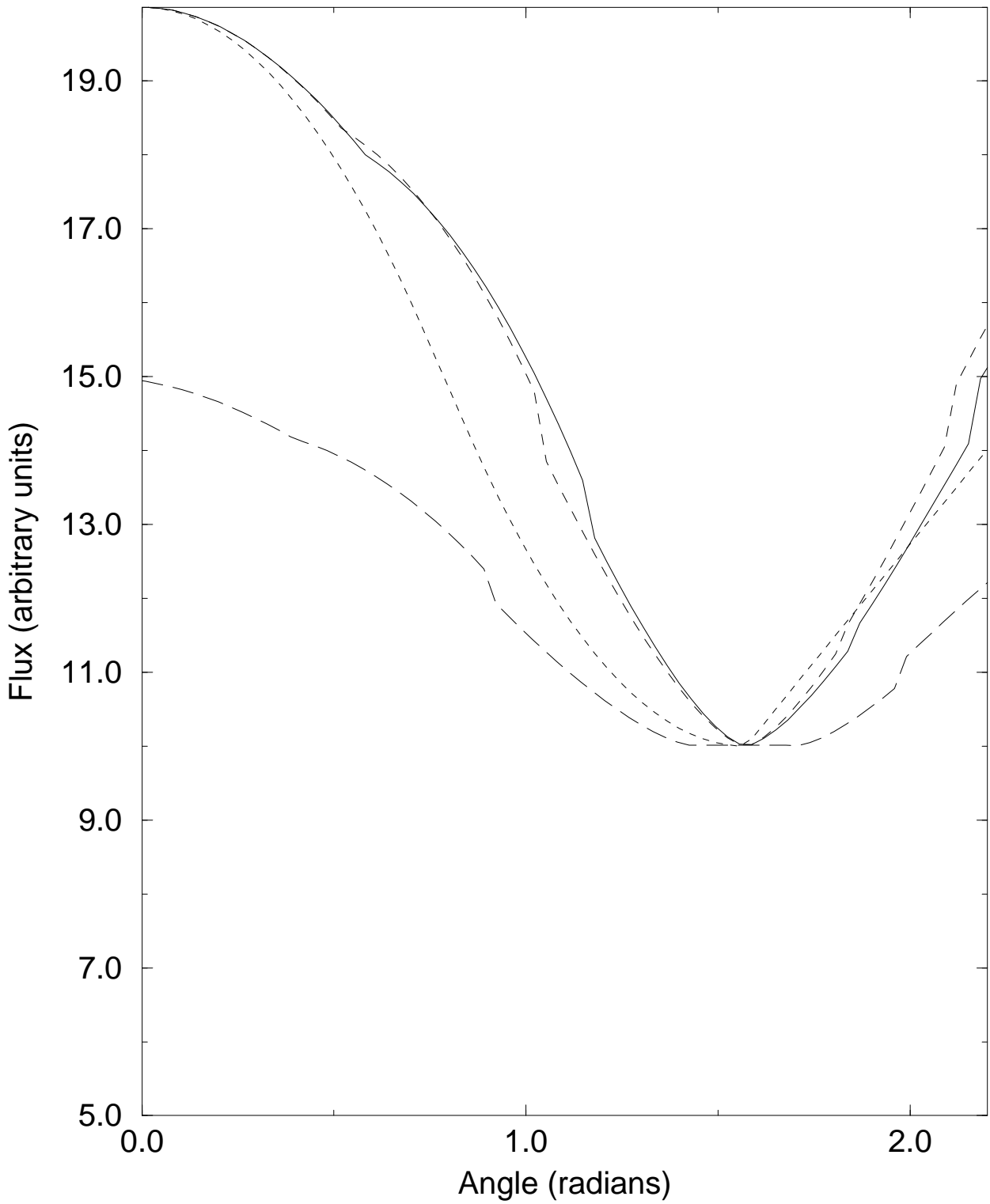
\begin{figure}[b]
\caption{An observer's view of the AGN in our model with covering 
factor of $f=0.5$. This is a
simplified form of the view from point G in Figure 2. The stepwise 
fashion which was chosen to describe
the absorber is for demonstrative purposes. See section \S 4 for details} 
\end{figure}

\begin{figure}[c]
\caption{The light curves near a dip that will be observed when the
source is eclipsed by the post-shock matter is shown for the softer
(dashed dotted) and the harder (solid curve) X-ray windows with $\phi=45^0$. 
The short dashed curve is for the case of a smaller Mach
surface opening angle of $\phi=30^0$, for the hard X-rays. The 
long-dashed curve
is for the simplified case in which a triangle with an opening 
angle $\phi=45^0$ eclipses a square source. In all cases
the maximum covering factor is $f=0.5$ (See section \S 4 for details).} 
\end{figure}

%

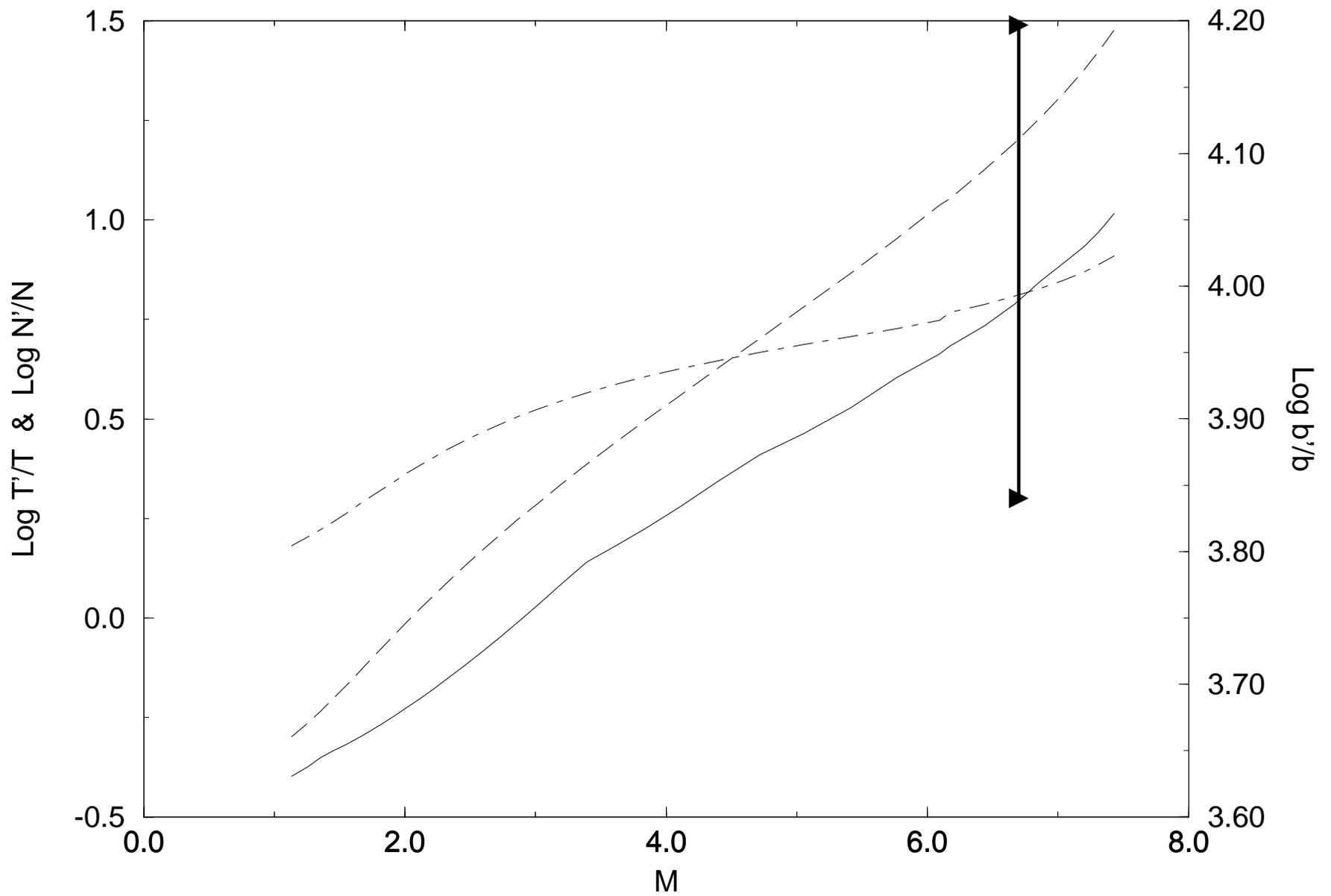
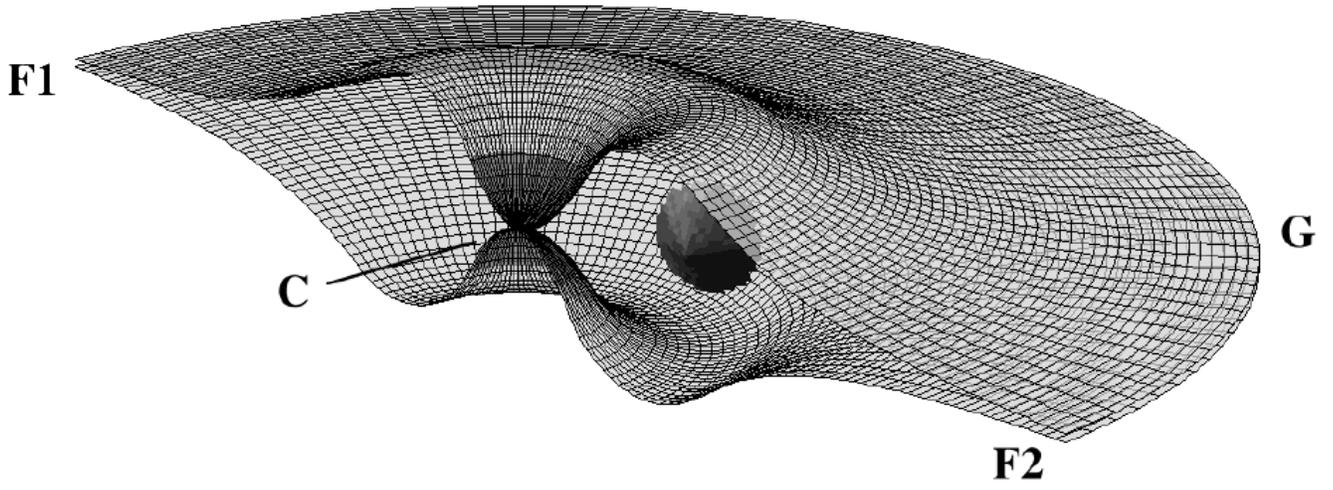
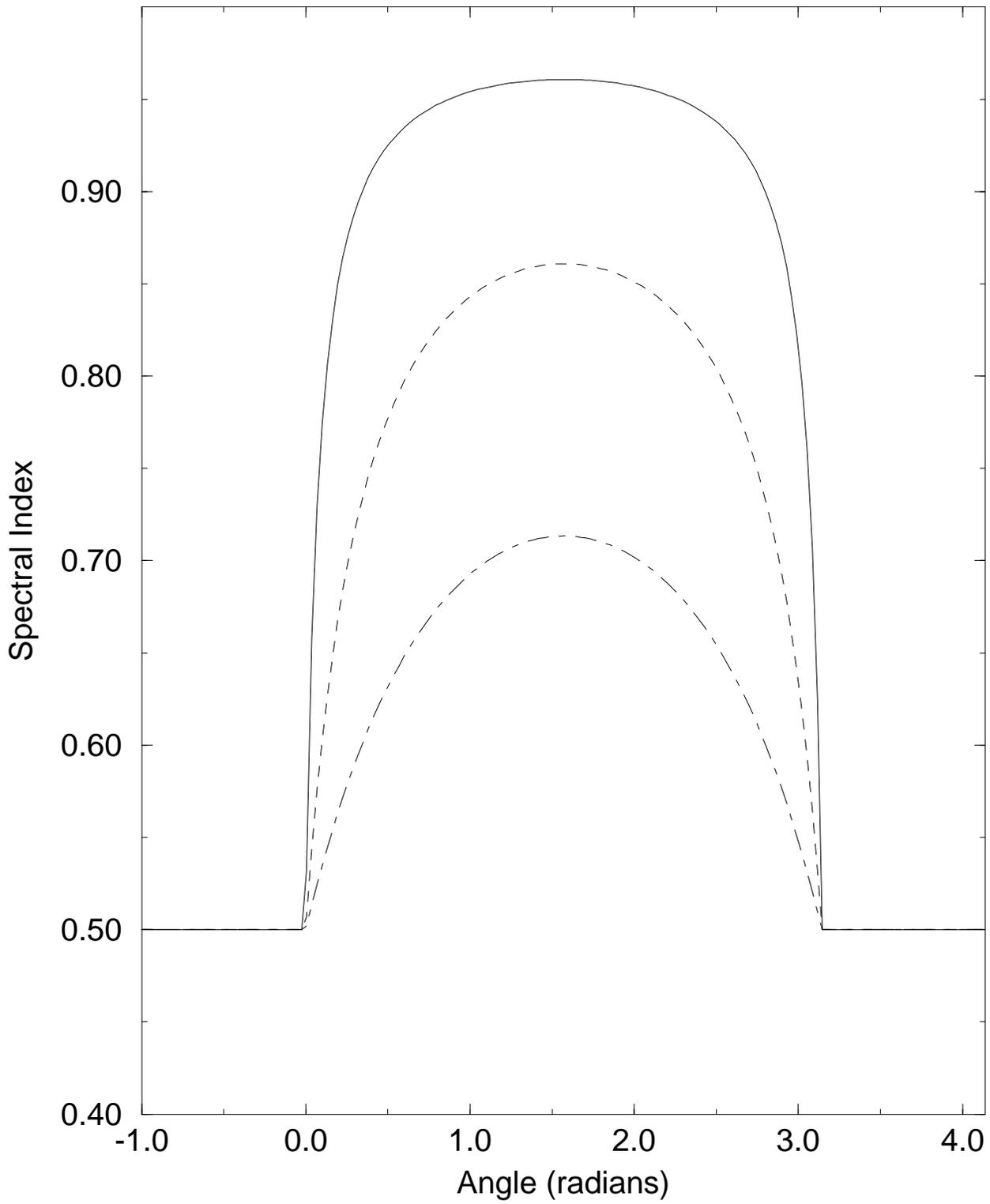
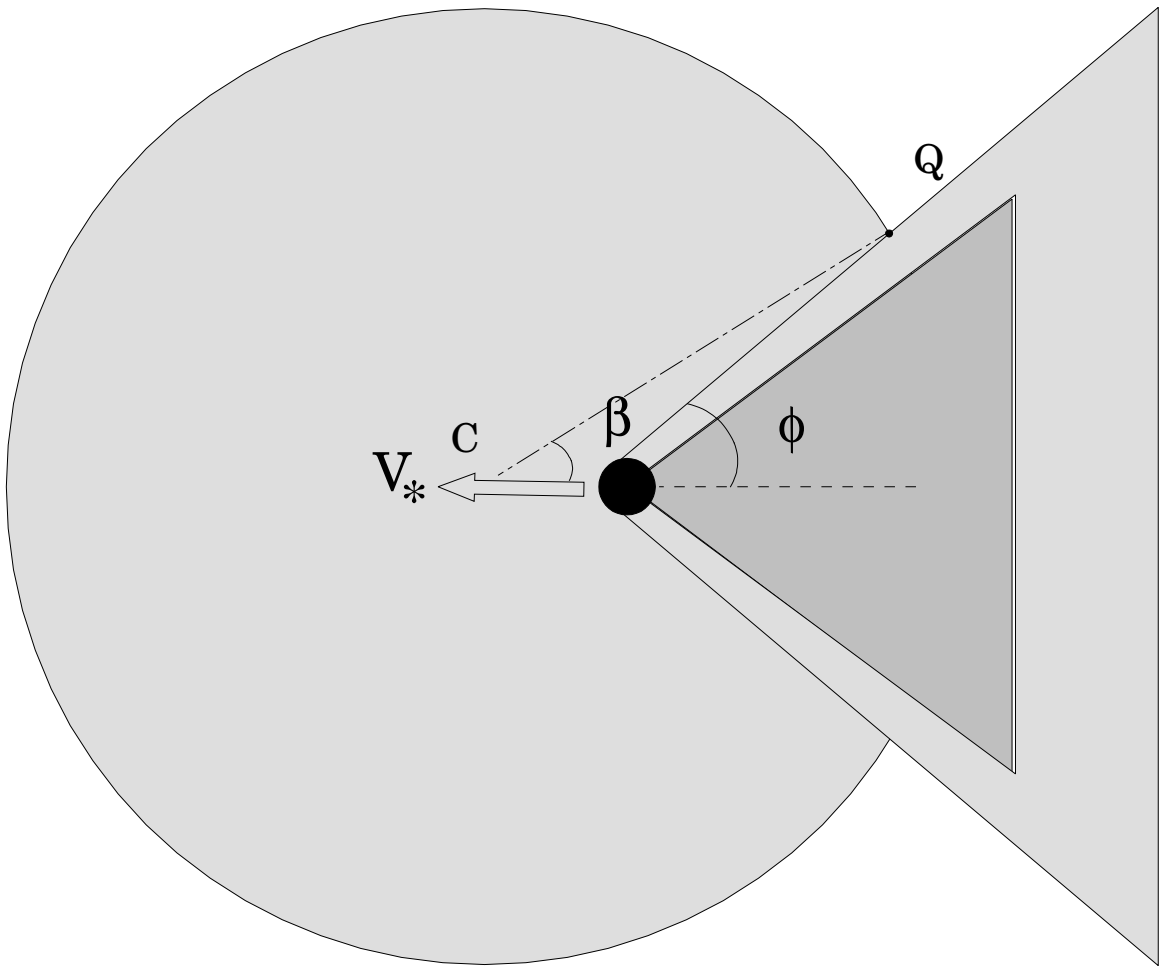
\end{document}